\documentclass{emulateapj}

\newcommand{\pwn}{3C~58}
\newcommand{\psr}{{PSR~J0205+6449}}
\newcommand{\chandra}{{\em Chandra}}
\newcommand{\spitzer}{{\em Spitzer}}

\usepackage{natbib,pstricks,mathptmx}

\slugcomment{To Appear in ApJ Letters}

\shorttitle{Spitzer Observations of 3C~58}
\shortauthors{Slane et al.}

\begin{document}

\title{The Infrared Detection of the Pulsar Wind Nebula in the Galactic
Supernova Remnant 3C~58}

\author{P. Slane\altaffilmark{1}, D. J. Helfand\altaffilmark{2},
S. P. Reynolds\altaffilmark{3}, B.~M. Gaensler\altaffilmark{4},
A. Lemiere\altaffilmark{1}, and Z. Wang\altaffilmark{5}}

\altaffiltext{1}{Harvard-Smithsonian Center for Astrophysics, 60 Garden
Street, Cambridge, MA 02138.}
\altaffiltext{2}{Columbia Astrophysics Laboratory, Columbia University, 550
West 120th Street, New York, NY 10027.}
\altaffiltext{3}{Department of Physics, North Carolina State University, Box
8202, Raleigh, NC 27695-8202.}
\altaffiltext{4}{School of Physics A29, The University of
Sydney, NSW 2006, Australia.}
\altaffiltext{5}{Department of Physics, McGill University, 3600 University
Street, Montreal, QC H3A 2T8, Canada.}

\begin{abstract}

We present infrared observations of 3C~58 with the {\em Spitzer Space
Telescope} and the Canada-France-Hawaii Telescope. Using the IRAC camera,
we have imaged the entire source resulting in clear detections of the
nebula at 3.6 and 4.5~$\mu m$. The derived flux values are consistent with
extrapolation of the X-ray spectrum to the infrared band, demonstrating
that any cooling break in the synchrotron spectrum must occur near the
soft X-ray band. We also detect the torus surrounding \psr, the 65~ms
pulsar that powers 3C~58. The torus spectrum requires a break between
the infrared and X-ray bands, and perhaps multiple breaks. This complex
spectrum, which is an imprint of the particles injected into the nebula,
has considerable consequences for the evolution of the broadband spectrum
of 3C~58.  We illustrate these effects and discuss the impact of these
observations on the modeling of broadband spectra of pulsar wind nebulae.

\end{abstract}


\section{Introduction}

\pwn\ is a flat-spectrum radio nebula ($\alpha \approx 0.1$, where $S_\nu
\propto \nu^{-\alpha}$) for which upper limits based on {\em IRAS}
observations indicate a spectral break between the radio and infrared
bands (Green \& Scheuer 1992).  X-ray observations reveal a nonthermal
spectrum from the nebula, with an average photon index $\Gamma = \alpha +
1 \sim 2.3$ (Torii et al. 2000) that varies with radius, becoming steeper
toward the outer regions of the nebula (Slane et al. 2004; hereafter S04).  
Subsequent
observations with the {\em Chandra X-ray Observatory} discovered \psr\
(Murray et al.  2002), the central pulsar in \pwn\ and one of the most
energetic pulsars known in the Galaxy, with a spin-down luminosity $\dot
E = 2.7 \times 10^{37}{\rm\ ergs\ s}^{-1}$. The pulsar powers a faint
jet and is surrounded by a toroidal structure apparently associated with
flows just downstream of the pulsar wind termination shock (S04).
The outskirts of the pulsar wind nebula (PWN) reveal faint thermal
emission associated with ejecta and/or ambient material swept up by \pwn\
(Bocchino et al. 2001; S04; Gotthelf et al. 2006).

\pwn\ has often been associated with SN 1181 (Stephenson \& Green
2005). However, the low break frequency would then suggest an unphysically
large magnetic field ($> 2.5 $mG) if interpreted as the result of
synchrotron losses. More recent investigations of the dynamics of
the system (Chevalier 2005), the radio expansion rate (Bietenholz 
2006), and the velocity of optical filaments (Rudie \& Fesen 2007)
argue convincingly that the age of the system is more like $\sim 2500$
years -- a value closer to the characteristic age of \psr\ ($\tau_c =
5.38 \times 10^3$~yr; Murray et al. 2002) -- but even at this age the
magnetic field in a synchrotron-loss scenario would be uncomfortably
large and inconsistent with the fact that the nonthermal X-ray emission
extends all the way to the radio boundary of the PWN (S04).
This discrepancy has prompted a number of explanations: 1) the pulsar
in \pwn\ underwent a rapid decline in its output at some early epoch
(Green \& Scheuer 1992); 2) the radiation spectrum reflects structure
in the spectrum of magnetic turbulence in the nebula (Fleishman \&
Bietenholz 2007); and 3) the low-frequency break is inherent in the
injection spectrum from the pulsar (Frail 1998).

The radial steepening of the X-ray spectrum in \pwn, while indicative
of synchrotron losses, is also difficult to understand in detail.
Assuming a power-law injection of particles into the nebula, the
spectral index varies with radius much more slowly than models for
an adiabatically-expanding nebula would predict (Reynolds 2003).
This problem of a low-frequency spectral break and a spectral index
that steepens slowly with radius is found in several other PWNe as
well, including G21.5$-$0.9 (Woltjer et al. 1997, Slane et al.
2000), making a more detailed investigation of the broadband
spectra of these nebulae important.

Here we report on observations of \pwn\ with the {\em Spitzer Space
Telescope (SST)} and Canada-France-Hawaii Telescope (CFHT). In \S2
we describe the observations and the data reduction procedures. In
\S3 we discuss the analysis of these data. A discussion and
interpretation of the broadband spectrum of \pwn\ is presented (\S4)
followed by our conclusions in (\S5).

\section{Observations and Data Reduction}

\pwn\ was observed by the {\em SST} on 18 January 2005 (Program ID
3647) using the Infrared Array Camera (IRAC).  A complete mapping
was carried out using a 54-position dither with the full array
readout, using both fields of view so as to obtain full coverage
of the PWN as well as a background region at all four wavelengths.
We used 100~s frametimes, yielding a minimum
exposure of 1200~s in the outer regions of the nebula, with increasing
overlap in the center leading to as much as 5400~s of total integration
time.

The IRAC data were processed and mosaiced using standard pipelines.
Images were obtained at 3.6, 4.5, 5.8, and 8 $\mu$m.  Our images reveal
emission from \pwn\ in both the 3.6 and 4.5~$\mu$m bands.  The nebula
is not seen above the very high foreground/background emission at 5.8
and 8 $\mu$m.  In Figure 1 we present the VLA image of \pwn\ (Reynolds
\& Aller 1985) along with the 4.5~$\mu$m image, with a single contour
representing the outer boundary of the radio emission.  The PWN is
clearly detected, with the infrared (IR) emission extending all the
way to the radio boundary in regions of the highest signal-to-noise.
The nebula is not detected at longer IRAC wavelengths where PAH emission
from Galactic dust dominates the field.

\begin{figure}
\plotone{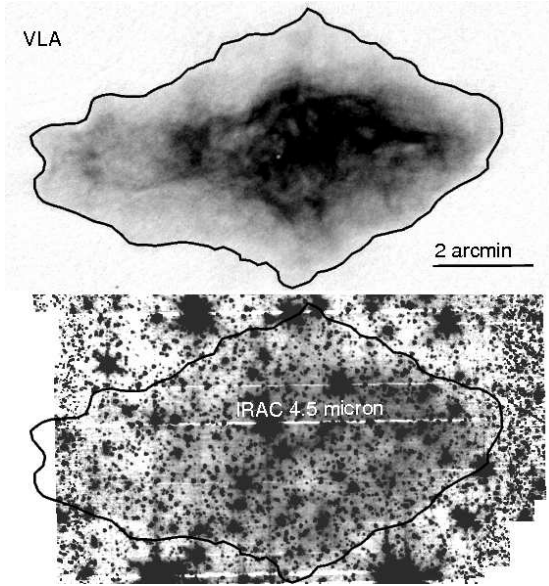}
\caption{
VLA and IRAC images of 3C58. The outermost radio contour is shown
on the IRAC (4.5~$\mu$m) image, where the PWN is clearly detected.
}
\end{figure}

While the emission in the shorter wavelength bands is severely
contaminated by stars, the overall morphology is clearly seen, and
the flux can be determined to within a factor of a few by scaling
the emission from numerous star-free regions, spread across the
PWN, to the entire size of the nebula. The overall morphology of the IR
emission from 3C~58 is strikingly similar to that seen in the radio
and X-ray bands (S04). The emission extends all the
way to the radio boundaries, indicating that no synchrotron loss
breaks occur at longer wavelengths. Some regions of
enhanced or diminished emission match well with those seen in the
other bands (notably the large cavity on the eastern side), suggesting
that we are observing primarily synchrotron radiation, although we
cannot yet rule out a component associated with shock-heated dust.
The derived flux does not demand a change in the
spectrum between the IR and X-ray bands, although the uncertainties
could accommodate a modest spectral break.

\chandra\ observations of \pwn\ reveal a central pulsar surrounded by an
extended torus (Slane et al. 2002). In Figure 2, we show the \chandra\
image of the pulsar, torus, and jet (left) alongside the IRAC 8~$\mu$m
image of the same region.  Remarkably, the IRAC image reveals emission
from the torus surrounding the pulsar. The source is extended on the
same spatial scales and is seen in all four IRAC bands. We have also
obtained a $J$-band image of this region from the Canadian Astronomy Data
Center. The image was made on 2002 August 16 with the KIR near-IR camera
at the Canada-France-Hawaii Telescope.  The torus emission observed in
IRAC is not detected in the $J$ image, in contrast with eight faint stars
in the field which are detected in both the $J$ and 3.6~$\mu$m images,
confirming that the extended emission has non-stellar colors.  The IRAC
observations represent only the second detection of a pulsar torus in
the IR band (the first being in the Crab Nebula -- Temim et al. 2006),
providing new constraints on the evolution of the particles as they
flow from the termination shock into the PWN (see Section 4). There is
no convincing evidence for emission from the pulsar jet.

\begin{figure}
\plotone{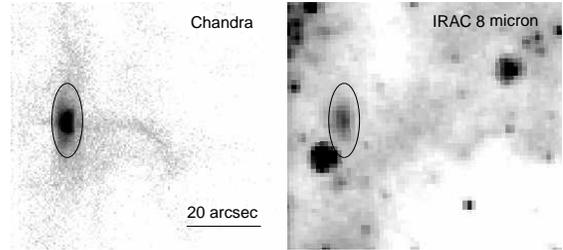}
\caption{Chandra image of \psr\ in 3C58, and its associated torus
(indicated by the ellipse) and jet (left) and the IRAC 8~$\mu$m image
of the same region (right). The torus is clearly detected here as well
as in the other three IRAC bands.}
\end{figure}

\section{Analysis and Interpretation}

We derived flux measurements for \pwn\ and its pulsar torus in each of
the IRAC bands. For the entire nebula, the diffuse emission is heavily
contaminated by stars (Figure 1). Our flux estimates are thus scaled
values based on measurements in small star-free regions. Because the
brightness of the diffuse emission is nonuniform, the uncertainty on
the scaled flux is large. Moreover, the background level also varies
considerably across the field. We have thus repeated the measurements
using several different background regions. The quoted uncertainties
are thus dominated by systematic uncertainties which we estimate at
$\sim 40\%$. Flux values for the detected regions are listed in Table
1, where we have corrected the observed flux values using extinction
corrections from Indebetouw et al. (2005).  As indicated in Figure 3,
where we have plotted the spectrum of 3C 58 from the radio band through
the X-ray band, the flux values measured by IRAC are consistent with
the extrapolation of the X-ray spectrum to the mid-IR band.

The morphology of the IR emission from 3C~58 is very similar to that seen
in the radio band, suggesting a synchrotron origin. In particular, there
is a good correspondence between discrete radio filaments and enhanced
emission in the IRAC images. Optical filaments in 3C~58 (van den Bergh
1978; S04; Rudie \& Fesen 2007), which presumably originate
from ambient gas overtaken by the expansion of the PWN, do not show a
good spatial correspondence with the radio or IR structures, suggesting
that the IR emission is not dominated by dust or line contributions. This
is similar to the results from {\sl Spitzer} observations of the Crab Nebula
(Temim et al. 2006), where emission in the IRAC band is also identified
primarily with synchrotron radiation.

For the pulsar torus, we have extracted the flux from a small ellipse
centered on the source, oriented north-south with semi-major and minor
axes of 7.8 and 4.9 arcsec (slightly smaller than that shown in Figure 2),
using a circular background region of radius 12.7~arcsec, centered on the
torus, with regions including the torus as well as faint stars removed.
To determine the flux uncertainties, we repeated the procedure using 
several different background
regions; we quote the difference between the highest and lowest flux
determinations using these background estimates.  The flux values are
presented in Table 1, and are plotted in Figure 3 along with the X-ray
spectrum (S04). The radio upper limit was determined from a
1.4 GHz VLA image (Reynolds \& Aller 1985) by adding to this image an
elliptical gaussian centered on the pulsar (with dimensions 28.0" x 7.9",
and with the long axis oriented north-south, as indicated in the \chandra\
image) and steadily increasing the flux until the simulated source was
readily detectable above the emission from the nebula.  The IRAC data for
the torus require a break in the spectrum between the X-ray and IR bands.

There is little question that torus emission is synchrotron in
nature; there is insufficient dust in the environment of the pulsar
termination shock to provide a shocked dust component to the emission.
We do not detect the pulsar jet (seen in X-rays -- Figure 2) in the
IRAC images, but this does not provide a significant constraint;
scaling $L_{IR}/L_x$ from the torus to the jet yields a predicted IR
flux that is well below the background levels in the IRAC images.


\begin{deluxetable}{lcccc}
\tablecolumns{5}
\tabletypesize{\scriptsize}
\tablewidth{0pc}
\tablecaption{IRAC Flux Densities}
\tablehead{
\colhead{Region} &
\colhead{3.6 $\micron$} &
\colhead{4.5 $\micron$} &
\colhead{5.8 $\micron$} &
\colhead{8 $\micron$} 
}
\startdata
PWN$^a$ & $2.8 \pm 1.2$ & $7.1 \pm 2.8$ & -- & --  \\
Torus$^b$ & $1.2 \pm 0.6$ & $1.6 \pm 0.8$ & $2.3 \pm 1.2$ & $3.4 \pm 1.7$  \\
\enddata
\tablecomments{
All flux densities are extinction-corrected. Dashes indicate non-detections.
Quoted errors correspond to systematic uncertainties.\\
a) Flux in units of $10^{-2}$~Jy \\
b) Flux in units of $10^{-4}$~Jy
}
\end{deluxetable}


\section{Discussion}

The broadband spectra of \pwn\ and its torus, shown in Figure 3, display
several distinct features. The overall spectrum from the PWN appears to
have two steepening breaks, one just beyond the radio band and another
somewhere in the IR band (indicated by dotted and dashed lines). The
torus spectrum also requires at least one steepening break between the
IR and X-ray bands (dashed line) -- a conclusion that has considerable
consequences for modeling of the PWN spectrum. If the spectral index
of the torus matches that of the PWN in the radio band (see discussion
below), the situation is even more complicated; the torus spectrum then
needs a steepening break between the radio and IR bands as well as two
breaks (one steepening and one flattening) between the IR and X-ray bands
(dotted lines).

The broadband spectrum of a PWN reflects both intrinsic and evolutionary
effects.  Some distribution of particles leaves the pulsar light
cylinder, and may evolve before reaching the termination shock where
it is thermalized and further modified.  Subsequent evolution in the
expanding nebula, including diffusion, convection, and energy losses in
the evolving nebular magnetic field, all affect the integrated spectrum
we observe.  For a power-law injection of particles from the pulsar,
a constant magnetic field in the nebula yields a power law synchrotron
spectrum with a break at a frequency
$\nu_b \approx 10^{21} B_{\mu \rm G}^{-3} t_3^{-2}{\rm\ Hz}$
(where $B_{\mu \rm G}$ is the magnetic field strength, in $\mu$G, and
$t_3$ is the age in units of $10^3$~yr)
above which the synchrotron cooling time of the radiating particles is
less than the age of the nebula. For an injection spectrum
$\frac{dN}{dE} = A E^{-s}$
where $N$ is the number of particles and $A$ is the normalization,
the radiated spectrum is
$S_\nu \propto \nu^{-\alpha}$
where $\alpha = (s-1)/2$ for $\nu < \nu_b$ and $\alpha = s/2$ for $\nu
> \nu_b$.  

\begin{figure}
\plotone{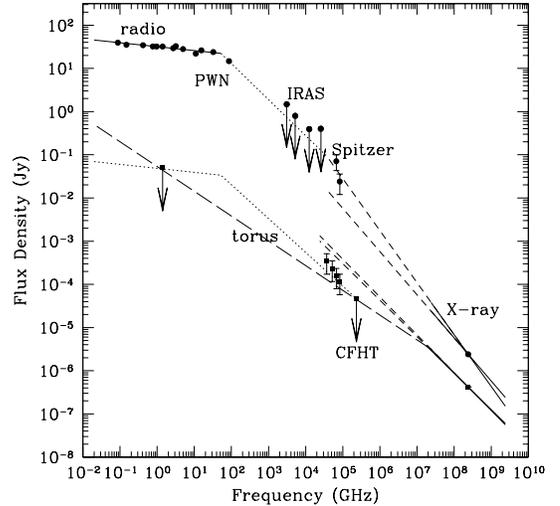}
\caption{The flux of all of 3C~58 (upper) and its torus (lower), plotted
from the radio to the X-ray band. While the torus is not detected in the
radio band, the IRAC data requires a flattening of the X-ray spectrum when
extrapolated back to the longer wavelength band. }
\end{figure}

The integrated spectrum of a PWN is, in fact, much more complicated than
a simple power law with a sharp synchrotron break (e.g., Reynolds \&
Chevalier 1984). Indeed, spectral breaks with $\Delta \alpha$ different
from 0.5 are possible in inhomogeneous models (e.g. Kennel \& Coroniti
1984, Reynolds \& Chevalier 1984), and are often observed in PWNe.
Variations in the magnetic field with time, to be expected as the PWN
expands, can also result in a broadening of the synchrotron break region.
In addition, the decline of the pulsar input power with time results in
modification of the photon spectrum with time.  Electron energies evolve
due both to adiabatic losses (where E scales inversely as the mean PWN
radius), and due to synchrotron losses (where electrons with energies
above an evolving cutoff energy all move to just below that energy).
Fossil or intrinsic breaks below the synchrotron break energy will
therefore evolve differently with time than the synchrotron break
frequency, while breaks above the synchrotron break frequency will
tend to be eliminated by synchrotron losses.  As the PWN expands and
the mean magnetic field weakens, the synchrotron-loss break will move
up in frequency, while any fossil breaks should move down (Woltjer et
al. 1997).  The broadband spectrum thus contains indications of the
nebula age, the pulsar input spectrum, and its evolution.

\begin{figure}
\plotone{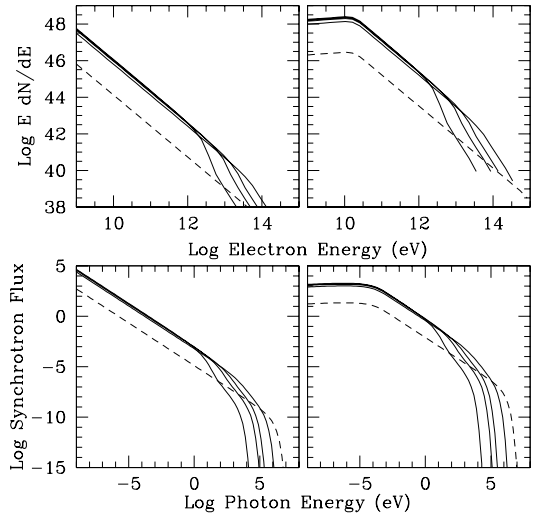}
\caption{
Simulated spectra for synchrotron emission from electrons in a PWN. Upper
panels show electron spectra for a power law input (a) and a broken
power law input (b), while panels (c) and (d) show the associated
synchrotron emission.  Dashed curves correspond to the injected particles
(integrated over 50 years), and solid curves represent the accumulation
of the particles (and associated emission) after 1000, 2000, 3000,
and 6000 years (from left to right). See text for description.
}
\end{figure}

While a complete modeling of the broadband spectrum of 3C~58, including
the effects of (asymmetric) adiabatic expansion, is beyond the scope of
this Letter, it is instructive to illustrate the effects of a break in
the particle injection spectrum.  In Figure 4, we plot synchrotron spectra
for an evolving population of electrons assuming two different forms for
the electron spectrum injected into a uniform magnetic field. In panel
(a), we show the electron spectrum for a power law injection, while panel
(b) shows a broken power law electron spectrum with a break energy at 60
GeV and indices of 1.1 and 3.3 below and above this break. The magnetic
field is $B = 10 \mu$G, and the upper energy limit on the electron
energy is set by the the condition that the electron gyroradius not
exceed the termination shock radius in 3C~58, $r_{ts} \approx 0.2$~pc
assuming a distance of 3.2~kpc.  The lower limit on the particle energy
is determined by the condition that the integrated electron energy not
exceed $\frac{1}{1+\sigma} \dot{E} t$, where $\dot{E}$ is the measured
spin-down loss rate for \psr, $t$ is the assumed age (see below), and
$\sigma = 0.001$ is the assumed magnetization parameter of the wind.
In each case, the dashed curve corresponds to the spectrum at
the time of injection (integrated over 50 years) and the solid curves
correspond to the built-up population after a period of 1000, 2000,
3000, and 6000 years. The spectral break at high energies is the result
of synchrotron losses.

In panels (c) and (d) we plot the associated synchrotron spectra
for these electron populations. It is evident that the injection
break in electron spectrum produces a low-energy break in the 
synchrotron spectrum whose break energy is constant in time, while
the energy at which the synchrotron-loss break appears decreases
with age. At the highest photon energies, the steep curvature of
the spectrum corresponds to the shape of the characteristic synchrotron
spectrum from electrons at the uppermost end of the particle spectrum.
This feature propagates downward in energy with increasing age as well.
Comparison with Figure 3 makes it clear that a single magnetic field
strength cannot reproduce the spectrum for both the nebula and
the torus; the torus requires a break between the X-ray and infrared
bands that is not seen in the spectrum of the nebula itself. Moreover,
if the low-frequency break observed in the PWN spectrum is an imprint
of the injection spectrum (as in the simple models in Figure 3), 
the torus spectrum then requires at least two breaks beyond the IR
band, one that flattens the spectrum and another that causes additional
steepening. The data alone are not sufficient to constrain the position
and magnitude of these breaks. More sophisticated modeling of the PWN
evolution, constrained by the observed emission from the torus region,
is required to understand the broadband emission.

\section{Conclusions}

We have presented \spitzer\ observations of \pwn\ that provide detections
of the nebula at 3.6 and 4.5 ${\mu}$m at flux values consistent with
an extrapolation of the X-ray spectrum into the mid-IR band. Moreover,
our detections of the torus surrounding \psr\ in each IRAC band, along
with upper limits from $J$-band observations, show that a spectral
break is required between the IR and X-ray bands, and that multiple
breaks are suggested.  These results show that the particles entering
the nebula through the torus do not have a simple power law spectrum,
and suggest that the low-frequency break in the large-scale nebula may be
associated with a break in the particle spectrum injected into the nebula.

To constrain realistic evolutionary models for the structure of 3C~58
that lead to its observed spectrum, it is crucial to obtain several
additional measurements for both the torus and the PWN. The position
of the low frequency break for the entire nebula is currently strongly
constrained by a single high-frequency measurement at $\sim 100$~GHz.
Confirmation of this result is crucial, as are measurements of the
flux in the sub-mm band.  For the torus, any improvements in the radio
upper limits will provide strong constraints. Just as important are
longer-wavelength measurements in the IR band, and deeper measurements in
the near-IR. The capabilities for each such measurement currently exist,
and hold promise for expanding our understanding of not only 3C~58,
but of all PWNe.

\acknowledgments
The work presented here was supported in part by {\sl Spitzer}
Grants JPL 1265776 (POS), JPL CIT 1264892 (DJH), and JPL RSA 1264893
(SPR), and as well as NASA Contract NAS8-39073 (POS).

\end{document}